\documentclass[aip,graphicx]{revtex4-1}
\usepackage{amsmath,amsfonts,amssymb}
\usepackage{booktabs}
\usepackage{graphicx}
\usepackage{epsfig}
\draft 

\begin{document}


\title{Enhancing multi-step quantum state tomography by PhaseLift} 



\author{Yiping Lu}

\affiliation{School of physics, Beijing Institute of Technology, Beijing 100081, China}

\author{Qing Zhao}
\email[]{qzhaoyuping@bit.edu.cn}
\affiliation{School of physics, Beijing Institute of Technology, Beijing 100081, China}


\date{\today}

\begin{abstract}
Multi-photon system has been studied by many groups, however the biggest challenge faced is the number of copies of an unknown state are limited and far from detecting quantum entanglement. The difficulty to prepare copies of the state is even more serious for the quantum state tomography. One possible way to solve this problem is to use adaptive quantum state tomography, which means to get a preliminary density matrix in the first step and revise it in the second step. In order to improve the performance of adaptive quantum state tomography, we develop a new distribution scheme of samples and extend it to three steps, that is to correct it once again based on the density matrix obtained in the traditional adaptive quantum state tomography. Our numerical results show that the mean square error of the reconstructed density matrix by our new method is improved to the level from $10^{-4}$ to $10^{-9}$ for several tested states. In addition, PhaseLift is also applied to reduce the required storage space of measurement operator.
\end{abstract}


\maketitle 

\section{Introduction}
\label{sec:level1}

Quantum entanglement is a necessary resource for quantum teleportation \cite{QuantumteleportationNature} and quantum computation \cite{QuantumcomputeNature}. It has been realized in photons, atoms and superconducting quantum circuits \cite{2,atomentangl,QuantumcomputeNature}. Certification of the entanglement of a state requires fidelity calculation. To further characterize the entanglement of a quantum state, an identical density matrix is an indispensable tool. Determination of this unique density matrix requires quantum measurements. Due to the characteristics of the destructiveness of the photon measurement, many copies of this unknown state are prepared to realize the quantum measurement. So far, it has been a challenge to prepare enough samples of the entangled state for detecting multi-photon entanglement in experiment \cite{2}. Therefore a scheme is required which consumes copies of a state as few as possible and thus making the prepared copies of the state are sufficient to confirm entanglement or reconstruct a density matrix. The whole process to reconstruct the density matrix is named as quantum state tomography (QST) \cite{James}. Specifically, the number of counts corresponding to different positive operator-valued measures (POVMs) is gained by repeatedly measuring the copies of a state. The ratio of the number of counts in one basis to the number of all counts detected in the same measurement setting is defined as a relative frequency, which can be obtained when the number of copies of the state is large enough. The relative frequencies are approximately regarded as probabilities. This measurement process follows Born's rule \cite{bORNRULE}. Then these probabilities can determine a density matrix by inverse transformation.

In QST, three main problems are encountered.

I, Several best measurement settings are required to be chosen for estimating the density matrix. Platonic solid measurements (the measurement bases are the vectors that connect the centers of the faces of the platonic solid and the center point in the Bloch sphere) give good performance of the reconstruction, and the overcomplete measurement sets can be commonly used to improve the accuracy of tomographic reconstruction \cite{MarkDdeBurgh}. However, these theoretical results are not easily accomplished in experiment, since the platonic solid measurements of overcomplete sets takes too much time in tomography.

II, Based on the measurement results (relative frequencies), it requires much time to reconstruct a multi-qubit density matrix. Hence, efficient algorithms are required to reconstruct a density matrix, the algorithms such as Maximum Likelihood (ML) estimation and the Least Squares (LS) method are commonly applied to QST \cite{Banaszek}. Recently, Compressive Sensing (CS) is also utilized to conduct tomography \cite{Davidgross2010}. Its performance is compared with ML in experiment \cite{Liuguofangkeda2012}, which shows that ML is better than CS. However, the conclusion may be taken with due care since the property of a density matrix is considered as the constraint of ML while it is not taken into account in the CS. Besides, a numerical result shows that CS outperforms LS when sampling rate is low \cite{AlexanderKron2014PRA}, while the opposite is true when sampling rate is high \cite{AlexanderKron2014PRA}. Based on the special characteristics of POVMs, PhaseLift is a good approach to save the storage space of POVMs, and has been also applied to solve a density matrix when the state is close to a pure state \cite{myarticle}.

III, The number of copies of an unknown state measured in each setting may not be equal. Identical copies of the arbitrary unknown quantum state are evenly distributed in each measurement setting in general tomography experiments. However, the required copies of a state increase sharply with the dimensions of the Hilbert space for the quantum states. Therefore a scheme needs to be explored to cut down a required number of  copies of the unknown state in QST. Recently, the adaptive QST is proposed to solve this problem \cite{D.H.Mahler,stefanLerch}. It divides the measurement into two steps. The first step is to get a preliminary density matrix and the second step is to correct it by measurements of the diagonal basis of the density matrix obtained previously \cite{D.H.Mahler}. It reduces the infidelity between the estimated and the true state from $O(1/\sqrt{N})$ to $O(1/N)$, where $N$ is the total number of samples of the unknown state \cite{D.H.Mahler}. This idea is further substantiated by one-qubit experiment \cite{thePaperOfGuoLaoshi}. The density matrix is reconstructed twice in the experiment.

In this paper, we improve the adaptive QST by reconstructing the density matrix three times by applying different distribution schemes of the copies of the unknown state in the second and third steps of measurements. Specifically, PhaseLift is applied to decrease the storage space of POVMs and we optimize the parameters by numerical simulation, such as the proportion of samples of an unknown state between the first step and the second step. A distribution scheme of copies of an unknown state for different settings is also developed, which helps to cut down the total number of copies of the state. The three-step adaptive QST is proposed to reduce the Mean Square Error (MSE).

The paper is organized as follows. In Section 2 and Section 3, PhaseLift and its connection with quantum state tomography are discussed. We are interested in applying the PhaseLift to adaptive quantum state tomography.
In Section 4, a new scheme is proposed to distribute different number of copies of a state into different settings. In Section 5, the numerical plots of the adaptive quantum state tomography via PhaseLift are given, and the three-step adaptive quantum state is compared with the conventional fixed quantum state tomography.

In the following part, various letters are used to represent different physical quantities. Let $N$ be the total number of copies of a state, $n$ denote the number of qubits, $R$ express the ratio of the number of copies of a state costed in the first step to the total number, $\rho_T$ represent a true unknown density matrix, $\rho_{E0}$ indicate the density matrix estimated in the first step and $\rho_E$ be the final estimated density matrix. Let $R_2$ be the ratio of the number of copies of a state costed in the third round of measurement to the total number.

\section{PhaseLift applied to Adaptive quantum state tomography}

PhaseLift is an approach to recover an unknown vector from several different measurements \cite{Candes33, E.J.CandesPHASELIFT2}. Let $\rho$ represent a general density matrix. When $\rho$ is a pure state, it can be decomposed as $PP^*$, where $P \in C^{2^n\times1}$ and the superscript $*$ indicates complex conjugate transpose. Let $\mu$ denote different settings, $\nu$ is used to distinguish different bases in a setting. Hence the POVMs corresponding to the $\nu$-th bases of the $\mu$-th setting can be described as $M_{\mu,\nu}$. Since $M_{\mu,\nu}$ also can be decomposed as $QQ^*$, and $Q\in C^{2^n\times1}$, then we have
\begin{eqnarray}
{\rm Tr}(\rho M_{\mu,\nu})=|\langle P,Q\rangle|^2.
\end{eqnarray}
From Ref.\cite{Candes33, E.J.CandesPHASELIFT2}, all the entries in $P$ can be recovered in the same phase difference. However, it has no effect on $\rho$ since the phase difference is eliminated in $PP^*$. Then the optimization model of the density matrix is constructed based upon the procedures given in Refs.\cite{James,Banaszek,21,Davidgross2010,Liuguofangkeda2012,Candes33,35,36}; and the noise is considered \cite{E.J.CandesPHASELIFT2,myarticle}.
\begin{eqnarray}
&& {\rm  Min} \sum_{\mu,\nu}|{\rm  Tr}(\rho M_{\mu,\nu})-f_{\mu,\nu}|  \nonumber\\
&& {\rm subject\ to}  \ \rho\geq0, {\rm  Tr}(\rho)=1\label{optidensitymatirx}
\end{eqnarray}
where $f_{\mu,\nu}$ is the frequency in the $\nu$th base of the $\mu$th setting and $\rho$ is the unknown density matrix, which is a positive semi-definite, and self-adjoint operator with trace equal to unity. This approach is applied to estimate the density matrix in the following work.

\section{Improved theory of adaptive quantum state tomography via PhaseLift}

The principle of adaptive quantum state tomography (AQST) is as follows: the copies of an unknown state $\rho_T$ is divided into two parts. One is used to estimate a density matrix $\rho_{E0}$, the other is used to adjust it. Specifically, a rough density matrix $\rho_{E0}$ is initially estimated by the uniform distribution of a part of the samples of $\rho_T$ in the all measurement settings. Then based on the characteristics of $\rho_{E0}$, only a part of settings are selected for the remaining measurements. From the measurement frequencies, a density matrix $\rho_{E}$ is estimated by considering the previous measurement frequencies \cite{D.H.Mahler}.

\begin{figure}
\includegraphics[width=12cm]{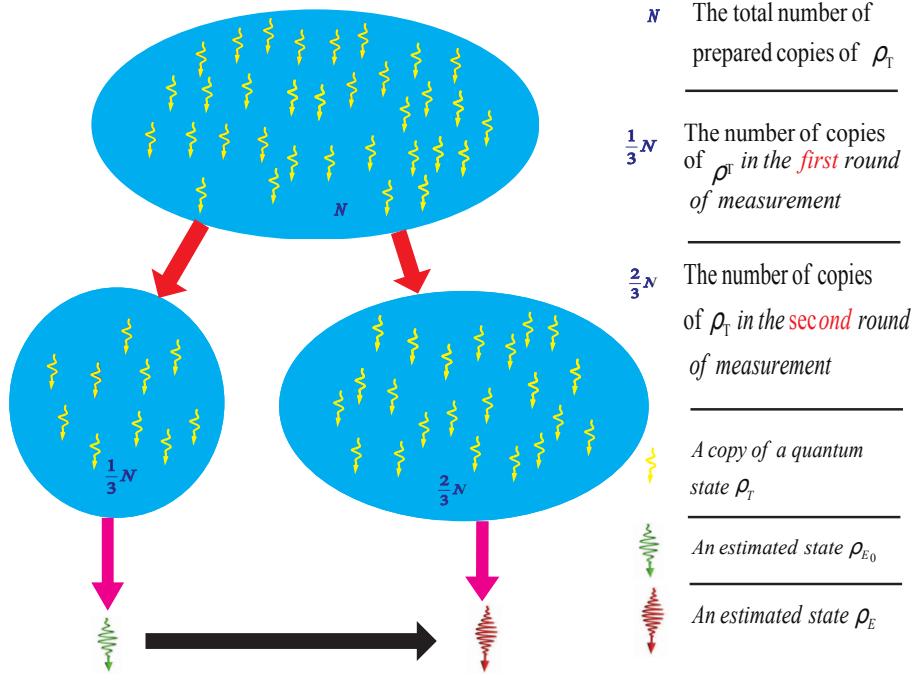}
\caption{\label{shiyituer} The process of adaptive QST. The copy of unknown state $\rho_{T}$ is represented by purple downward spiral line. Traditional adaptive QST is to split the total number of copies of unknown state, N, into two part. One costs 1/3 of the total copies of the state, the other costs 2/3 of the total copies of the state.  From the first part, a preliminary density matrix $\rho_{E0}$ (represented by green downward spiral line) is obtained. Then based on the measurement results from the other part, $\rho_{E}$ represented by red downward spiral line is obtained.}
\end{figure}

We improve AQST by proposing a scheme to select the best settings and applying PhaseLift to obtain the density matrix. Firstly Pauli measurement is applied in our simulation. Hence the total number of settings is $3^n$ for $n$-qubit system, and $\mu$ is an integer between $1$ and $3^n$. The superscript $(1)$ indicates the first round of measurements. Then the frequency in the $\nu$-th base of the $\mu$-th setting in the first round of measurements can be described as $f_{\mu,\nu}^{(1)}$, which is obtained by measuring copies of $\rho_T$ with the number of $N\cdot R/3^n$ on the POVMs. Then a roughly estimated density matrix $\rho_{E0}$ can be given via PhaseLift by inputting $f_{\mu,\nu}^{(1)}$. The new scheme is applied to select measurement settings and to properly assign the remaining copies of $\rho_T$ to the selected settings in the second round of measurements. Based on the count distributions, the relative frequencies $f_{\mu,\nu}^{(2)}$ are gained in the chosen settings. By forming linear combinations of the former frequencies $f_{\mu,\nu}^{(1)}$, the new relative frequencies $f_{\mu,\nu}^{(2)}$, $f_{\mu,\nu}^{(3)}$ are obtained. Finally, a new density matrix $\rho_E$ can be obtained by inputting these new frequencies $f_{\mu,\nu}^{(3)}$ into PhaseLift, as shown in Fig.\ref{shiyituer}.

\section{Selection scheme of setting}
After obtaining $\rho_{E0}$ in the first round of measurements on the Pauli bases, the new mechanism for getting a more accurate density matrix $\rho_{E}$ is described in detail in this section. Its purpose is to select the distribution of the number of copies of an unknown state $\rho_T$ in the different settings for the second round of measurements. Since $N\cdot R$ copies of $\rho_T$ are consumed in the first round of measurements, the remaining copies of $\rho_T$ have the number $N\cdot (1-R)$ which are used to  reduce the error of each entry of the density matrix $\rho_{E0}$ as much as possible. Since the error caused by an entry with large modulus is also big \cite{D.H.Mahler}, a large number of samples of $\rho_T$ are required to reduce the errors of these entries. Then Mean Square Error (MSE) between $\rho_T$ and $\rho_E$ can be relatively small. In other words, a portion of measurement settings is selected to ensure that their bases are enough to uniquely decompose the density matrix. If the coefficient of a basis in the decomposition is relatively large, it means that the basis for determining the unknown density matrix is more important or it has a much larger contribution to cause the deviation of the desired density matrix \cite{PaulimeasuremntFidelity}. Hence, more than the average number of copies of the unknown state $\rho_T$ is expected to be distributed on these bases. The coefficients are obtained by decomposing the roughly estimated density matrix $\rho_{E0}$. The summation of coefficients of the bases in the same setting determines the number of samples of $\rho_T$ used for each setting in the second round of measurement. The coefficients and the number of samples are chosen to be linear proportional. Specifically, let $\overrightarrow{S}$ be such a vector that each entry $S_{\mu,\nu}$ of $\overrightarrow{S}$ is the coefficient of the $\nu$th basis of the $\mu$th settings ($M_{\mu,\nu}$), then the summation of coefficient in the same setting is
$S_{\mu}=\sum_{\nu} S_{\mu,\nu}$.

 The purpose of this mechanism is to find out the measurement settings as small as possible and the bases belong to these settings are sufficient to uniquely decompose the density matrix $\rho$. Since $\rho$ is a positive semidefinite matrix, the following model is proposed to select settings based on the above principle.
\begin{eqnarray}
{\rm Min}\ ||\overrightarrow{S}||_1,\ {\rm subject \ to}\ \overrightarrow{S}\cdot \overrightarrow{M}=\rho, \
\rho\geq 0,\label{12}
\end{eqnarray}
where
\begin{eqnarray}
&& \overrightarrow{S}=(S_{1,1},S_{1,2},\cdots,S_{1,2^n},S_{2,1},S_{2,2},\cdots,S_{\mu,\nu},\nonumber\\
&&\cdots,S_{3^n,2^n})^T
\end{eqnarray}
and $\overrightarrow{M}$ is a vector whose elements are the bases of different measurement settings. Usually these bases can be either the bases of all different Pauli operators or mutually unbiased bases \cite{mub}. Here the bases of all different Pauli operators are applied for numerical tests since they are much easier to utilize in experiment \cite{JAMES}. Then eigen-vectors can be calculated for each tensor product of Pauli operators. Specifically, the local Pauli operator $\sigma_x$, $\sigma_y$, $\sigma_z$ are applied. Let $|0_x\rangle$ and $|1_x\rangle$ denote the two eigen-vectors of $\sigma_x$, $|0_y\rangle$ and $|1_y\rangle$ for $\sigma_y$, and $|0_z\rangle$ and $|1_z\rangle$ for $\sigma_z$. Then the elements of $\overrightarrow{M}$ for the two-qubit case are as follows:
\begin{eqnarray}
&& \overrightarrow{M}=(|0_x0_x\rangle\langle0_x0_x|,|0_x1_x\rangle\langle0_x1_x|,|1_x0_x\rangle\langle1_x0_x|, \nonumber\\
&& |1_x1_x\rangle\langle1_x1_x|, |0_x0_y\rangle\langle0_x0_y|, |0_x1_y\rangle\langle0_x1_y|,|1_x0_y\rangle\langle1_x0_y|,\nonumber\\
&&|1_x1_y\rangle\langle1_x1_y|,\cdots, |0_z0_z\rangle\langle0_z0_z|,|0_z1_z\rangle\langle0_z1_z|,\nonumber\\
&&|1_z0_z\rangle\langle1_z0_z|,|1_z1_z\rangle\langle1_z1_z|).
\end{eqnarray}
 To facilitate the solution, the model of Eq. (\ref{12}) is transformed into the following form:
\begin{eqnarray}
{\rm Min}\ ||\overrightarrow{S}||_1,\ {\rm subject \ to}\ ||\overrightarrow{S}\cdot \overrightarrow{M}-\rho||_2\leq \epsilon, \
\rho\geq 0,\label{selectioner}
\end{eqnarray}
where $\epsilon$ is a maximum allowed error and set to 0.00001 in the following numerical simulation. Since $\overrightarrow{S}$ can be obtained from the model of Eq.(\ref{selectioner}), then the remaining copies of the unknown state $\rho_T$ with the number of $N\cdot(1-R)$ are allocated according to $\overrightarrow{S}$. The rule to allocate the copies of the unknown state $\rho_T$ is based on the summation of coefficients of the bases belonging to the same setting. More copies of the unknown state $\rho_T$ need to be measured when the value is large. In other words, the following rule is proposed for copies' distribution.
\begin{eqnarray}
\frac{S_{\mu}}{||\overrightarrow{S}||_1}=\frac{\sum_\nu S_{\mu,\nu}}{||\overrightarrow{S}||_1}=\frac{N_\mu}{\sum N_\mu},
\end{eqnarray}
where $N_\mu$ denotes the number of copies of the unknown state $\rho_T$ for the $\mu$-th setting. Relative frequency of the second round of measurement $f_{\mu,\nu}^{(2)}$ can be obtained after measuring samples $\rho_T$ with the number of $N\cdot(1-R)\cdot N_\mu/\sum N_\mu$ on the POVMs $M_{\mu,\nu}$. The superscript $(2)$ in $f_{\mu,\nu}^{(2)}$ represents the second round of measurement.

Sparse property of the density matrix is taken as a priori. Since the frequencies corresponding to cases where POVMs measure the zero entries of a sparse density matrix or their linear combination are zero if there is no noise. However, some noise always exists, therefore, the frequency for this kind of POVMs is a little larger than zero in practical measurements. Because the summation of frequencies for all the bases in the same setting are always equal to one, the POVMs measuring the non-zero entries of a sparse density matrix may correspond to a relatively larger frequency than the detected one ($f_{\mu,\nu}^{(2)}$). Therefore, $f_{\mu,\nu}^{(2)}$ corresponding to the large coefficient $S_\mu$ of the setting is adjusted to a slightly larger value by a factor $3^nN_\mu/(\sum N_\mu)$, then the relative frequency $f_{\mu,\nu}^{(3)}$ takes the form
\begin{eqnarray}
f_{\mu,\nu}^{(3)}=R\cdot f_{\mu,\nu}^{(1)}+(1-R)\cdot f_{\mu,\nu}^{(2)}\frac{3^nN_\mu}{\sum N_\mu}.\label{FUM}
\end{eqnarray}
Then $f_{\mu,\nu}^{(3)}$ is used to get a more precise density matrix $\rho_{E}$ by applying PhaseLift once again, as shown in Fig.\ref{shiyituer}.

\begin{figure}[H]
\begin{center}
\includegraphics[scale=0.7]{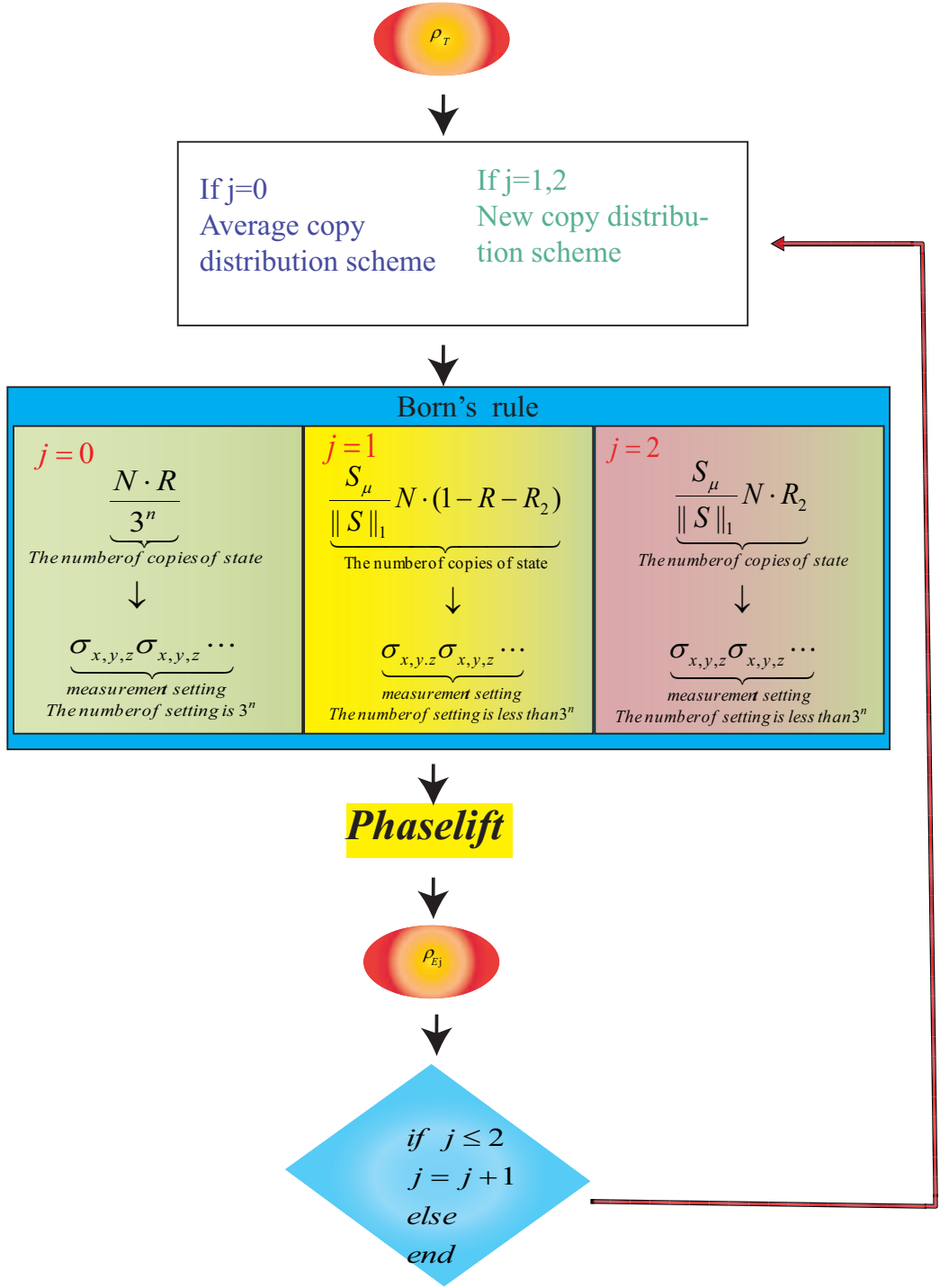}
\caption{\label{Flowcharter} The flow chart of three-step AQST.}
\end{center}
\end{figure}

\begin{figure}[H]
\begin{center}
\includegraphics[width=10cm]{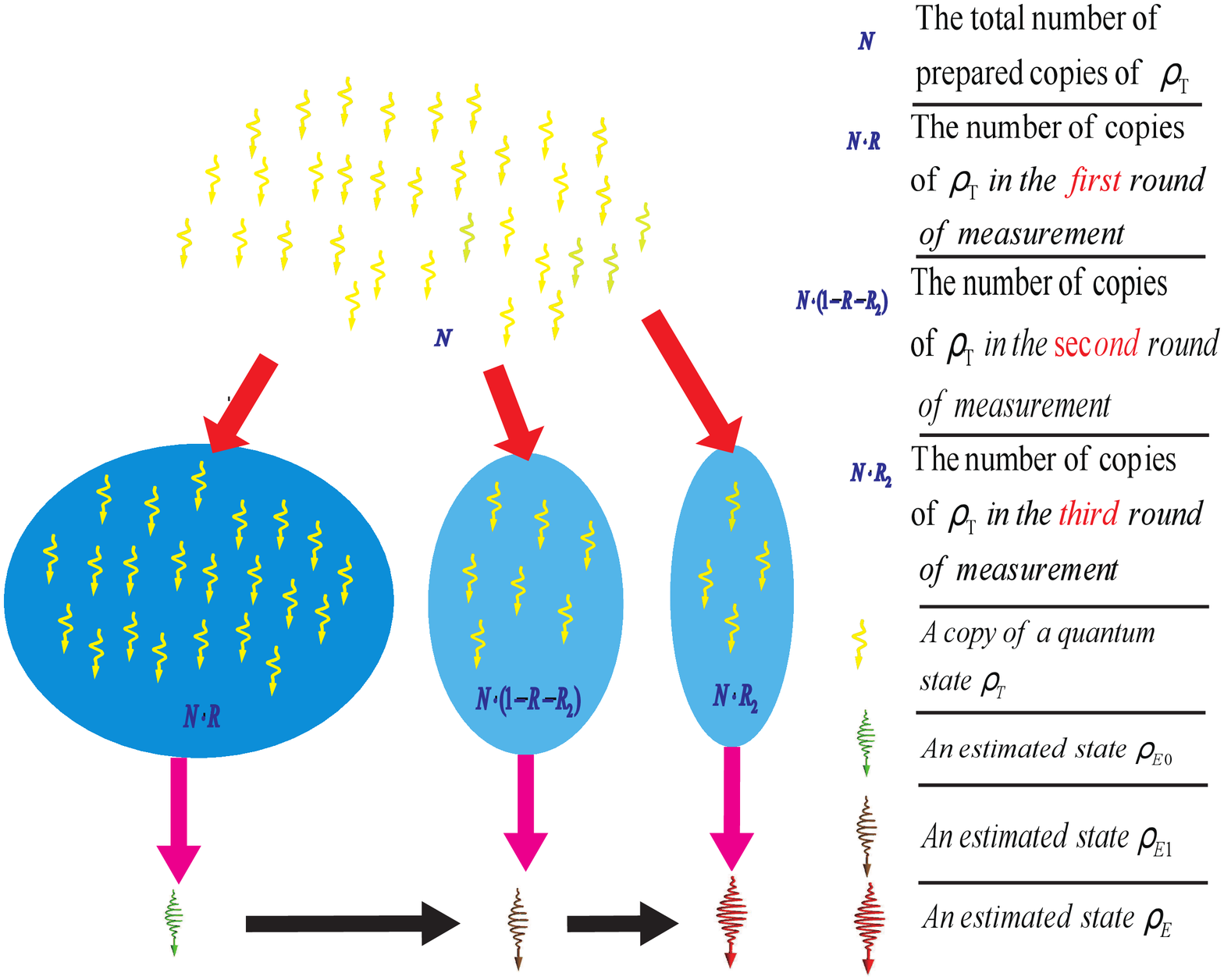}
\caption{\label{shiyituerThreestep} The process of three-step AQST. The copy of unknown state $\rho_{T}$ is represented by purple downward spiral line. Three-step QST is to split the total number of copies of unknown state, N, into three part. One costs $N\cdot R$ of copies of the state, the second costs $N\cdot(1-R-R_2)$ of copies of the state, and the last part costs $N\cdot R_2$. From the first part, a rough density matrix $\rho_{E0}$ (represented by green downward spiral line) is obtained. Combined with the results measured by the second part of copies of the state, a better density matrix $\rho_{E1}$ represented by brown downward spiral line is obtained. Then based on the $\rho_{E1}$ and the measurement results from the last part, $\rho_{E}$ represented by red downward spiral line is obtained.}
\end{center}
\end{figure}

To further improve the reconstructed density matrix, the three-step AQST is developed. The difference between two-step and three-step AQST is that the density matrix $\rho_{E0}$ is revised twice in three-step AQST while the density matrix $\rho_{E0}$ is amended only once in two-step AQST by using the same number of copies of unknown state $\rho_{T}$. In other words, the copies of unknown state $\rho_{T}$ for the second step of measurement in two-step AQST is divided into two parts in three-step AQST, one is applied to get a little precise density matrix $\rho_{E1}$ than $\rho_{E0}$. Then based on the $\rho_{E1}$, the other part of copies of $\rho_{T}$ is measured to further revise the $\rho_{E1}$ to get a more precise $\rho_{E}$, as shown in Fig.\ref{Flowcharter} and Fig.\ref{shiyituerThreestep}. The $f_{\mu,\nu}^{(4)}$ can be calculated by the density matrix $\rho_{E1}$:
\begin{eqnarray}
f_{\mu,\nu}^{(4)}={\rm Tr}(M_{\mu,\nu}\rho_{E1}).
\end{eqnarray}
Based on the same idea of taking sparsity as a priori, the frequency gets modified again,
\begin{eqnarray}
&&f_{\mu,\nu}^{(5)}=R \cdot f_{\mu,\nu}^{(1)}+R_2 \cdot f_{\mu,\nu}^{(2)}\frac{3^nN_\mu}{\sum N_\mu}\nonumber\\
&&+(1-R-R_2) \cdot f_{\mu,\nu}^{(4)}\cdot \frac{3^nN_\mu}{\sum N_\mu}
\end{eqnarray}
then $f_{\mu,\nu}^{(5)}$ is applied to obtain $\rho_{E}$ by PhaseLift, therefore $\rho_{E}$ is the finial density matrix.

\section{Numerical simulation results}

\begin{figure}[H]
\begin{center}
\includegraphics[width=10cm]{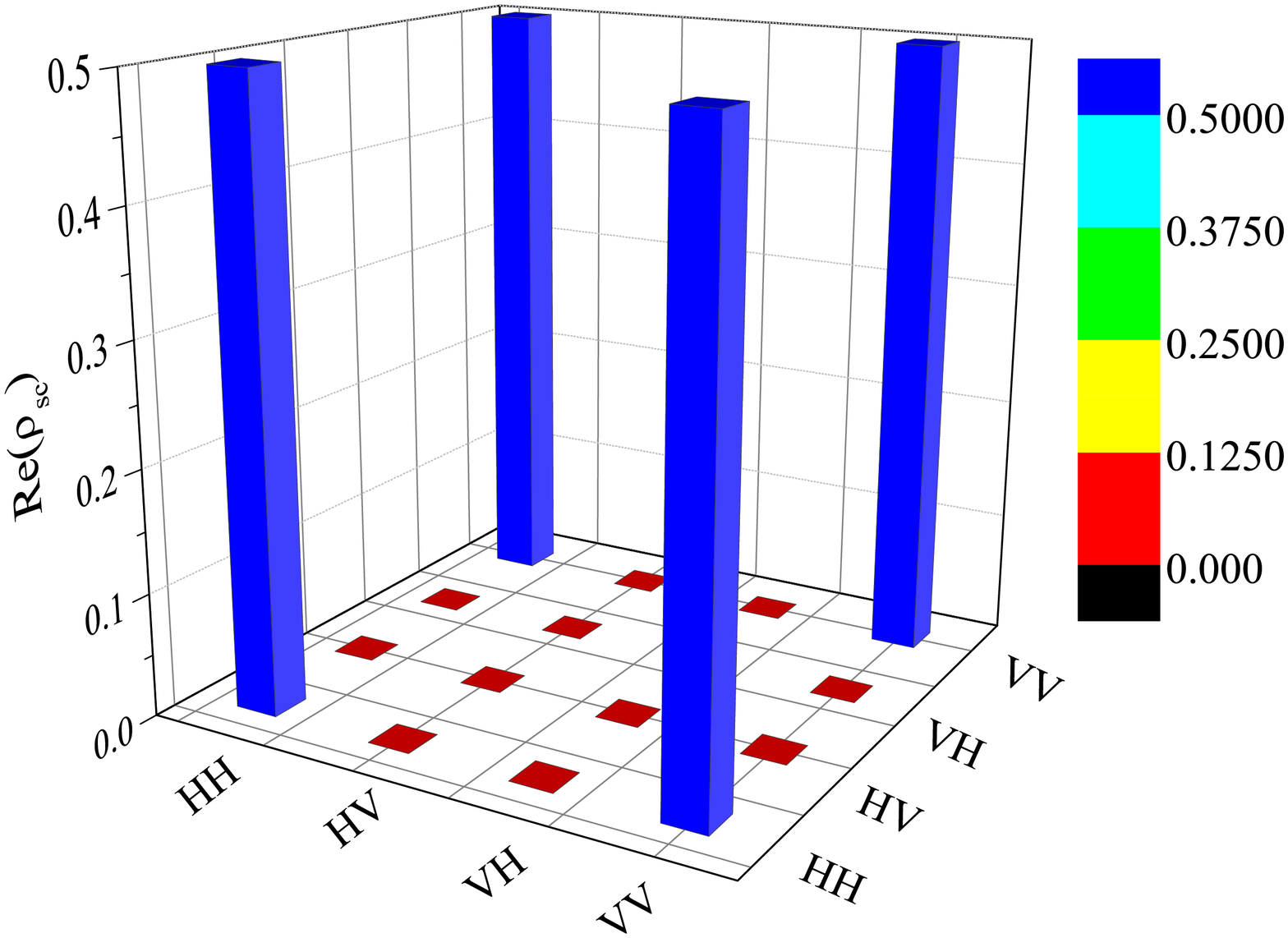}
\caption{\label{Sc}Density matrix of two-qubit SC state.}
\end{center}
\end{figure}

\begin{figure}[H]
\begin{center}
\includegraphics[width=10cm]{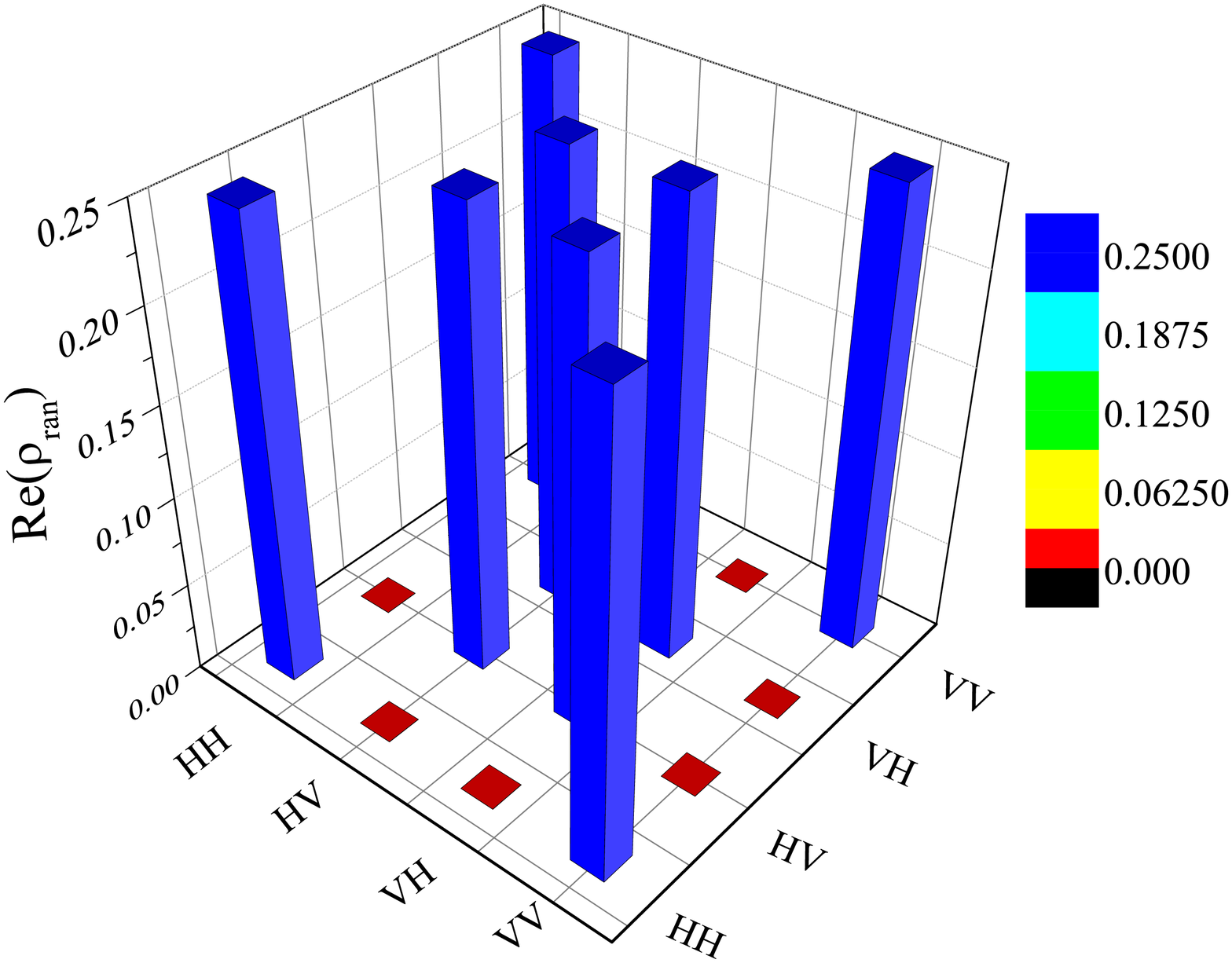}
\caption{\label{G9} Density matrix randomly generated.}
\end{center}
\end{figure}

 Several states are selected for numerical simulations. These quantum states include Schr\"{o}dinger's Cat (SC) state, as shown in Fig.\ref{Sc}, NOON state \cite{noon}, W state \cite{W} and a quantum state randomly generated but satisfying the properties of the density matrix, as shown in Fig.\ref{G9}.


\begin{figure}
    \centering
    \includegraphics[width=10cm]{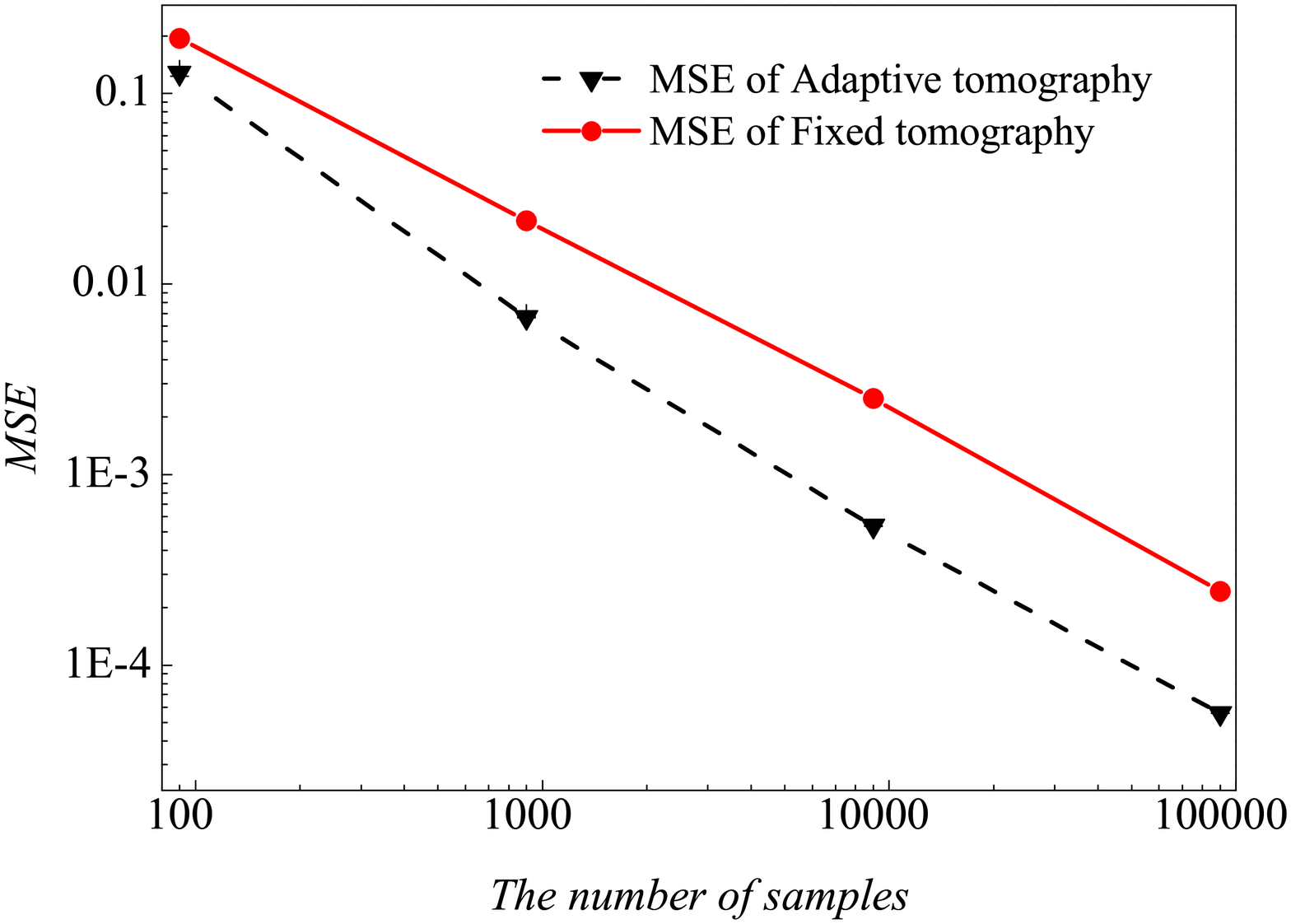}
     \caption{\label{fuc} MSE under the different number of samples of an unknown state when $R$ is set to 1/2.}
\end{figure}

 Two-step AQST is firstly tested. Mean Square Error (MSE) between $\rho_T$ and $\rho_E$ is compared in AQST and fixed QST. The two cases are indicated by "$MSE$ $of$ $Adaptive$ $tomography$" and "$MSE$ $of$ $Fixed$ $tomography$" respectively. ``Fixed" is used because each setting is measured only once with the same amount of copies of the unknown state. Fixed QST is the standard quantum state tomography (SQST), which means the density matrix is estimated directly using the PhaseLift after the measurement.  It is observed that MSE between the estimated and true density matrices decreases when the copies of $\rho_T$ increase. They have approximately a linear relationship when exponential coordinates are used for both MSE and the number of samples of $\rho_T$, as shown in Fig.\ref{fuc}.

\begin{figure}\centering
    \includegraphics[width=10cm]{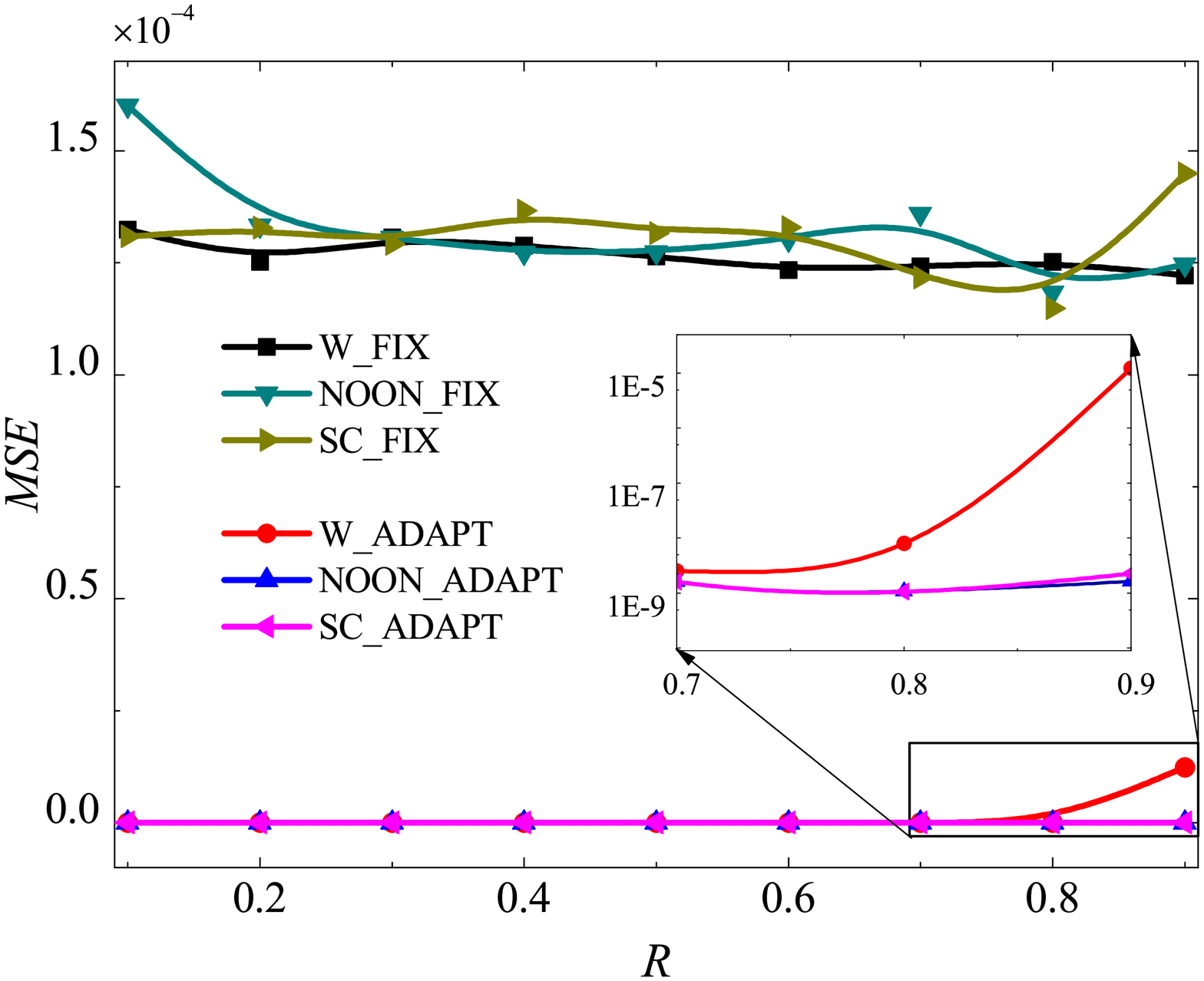}
    \vspace{-3em}
    \caption{\label{PURESCNOONstatesThreeW} Mean Square Error (MSE) under different $R$s. ``$SC$$\underline{\hbox to 2mm{}}$$FIX$" denotes the fixed quantum state tomography for SC state; ``$SC$$\underline{\hbox to 2mm{}}$$ADAPT$" denotes the AQST for SC state; ``$NOON$$\underline{\hbox to 2mm{}}$$FIX$" denotes the fixed quantum state tomography for NOON state; ``$NOON$$\underline{\hbox to 2mm{}}$$ADAPT$" denotes the AQST for NOON state;  ``$W$$\underline{\hbox to 2mm{}}$$FIX$" denotes the general quantum state tomography for W state; And ``$W$$\underline{\hbox to 2mm{}}$$ADAPT$" denotes the AQST for W state.}
\end{figure}

Different quantum states are applied to test the two-step AQST. Let $R$ denote the ratio of the number of samples of a state applied in the first step to the total number of samples of the state. A pure two-qubit SC state is chosen as the true density matrix $\rho_T$. Quantum state is reconstructed from the same total samples of $\rho_T$ (90000) when they are distributed according to different $R$'s. Then MSE between the final reconstructed density matrix $\rho_E$ and the $\rho_T$ is calculated. $R$ in each case is repeated for 50 times. It is observed that AQST produces much better results ($10^{-9}$ of MSE) than fixed QST ($10^{-4}$ of MSE). The variance of ratio $R$ does not have a marked impact on the change of error of the reconstruction. When the true density matrix $\rho_T$ is a pure two-qubit NOON state, the same process is repeated as the two-qubit SC state except each MSE is gained by 100 times of MSE for one $R$. The same results are observed. The optimal $R$ for AQST is nearly 1/2. Besides, three-qubit W state is also tested. The MSE is calculated in different $R$ when $270000\cdot R$ copies of W state are applied in the first step and $270000\cdot(1-R)$ copies of W state are applied in the second step, as shown in Fig.\ref{PURESCNOONstatesThreeW}. So the two-step AQST outperforms the fixed QST for all the states tested above.

\begin{figure}
    \centering
    \includegraphics[width=10cm]{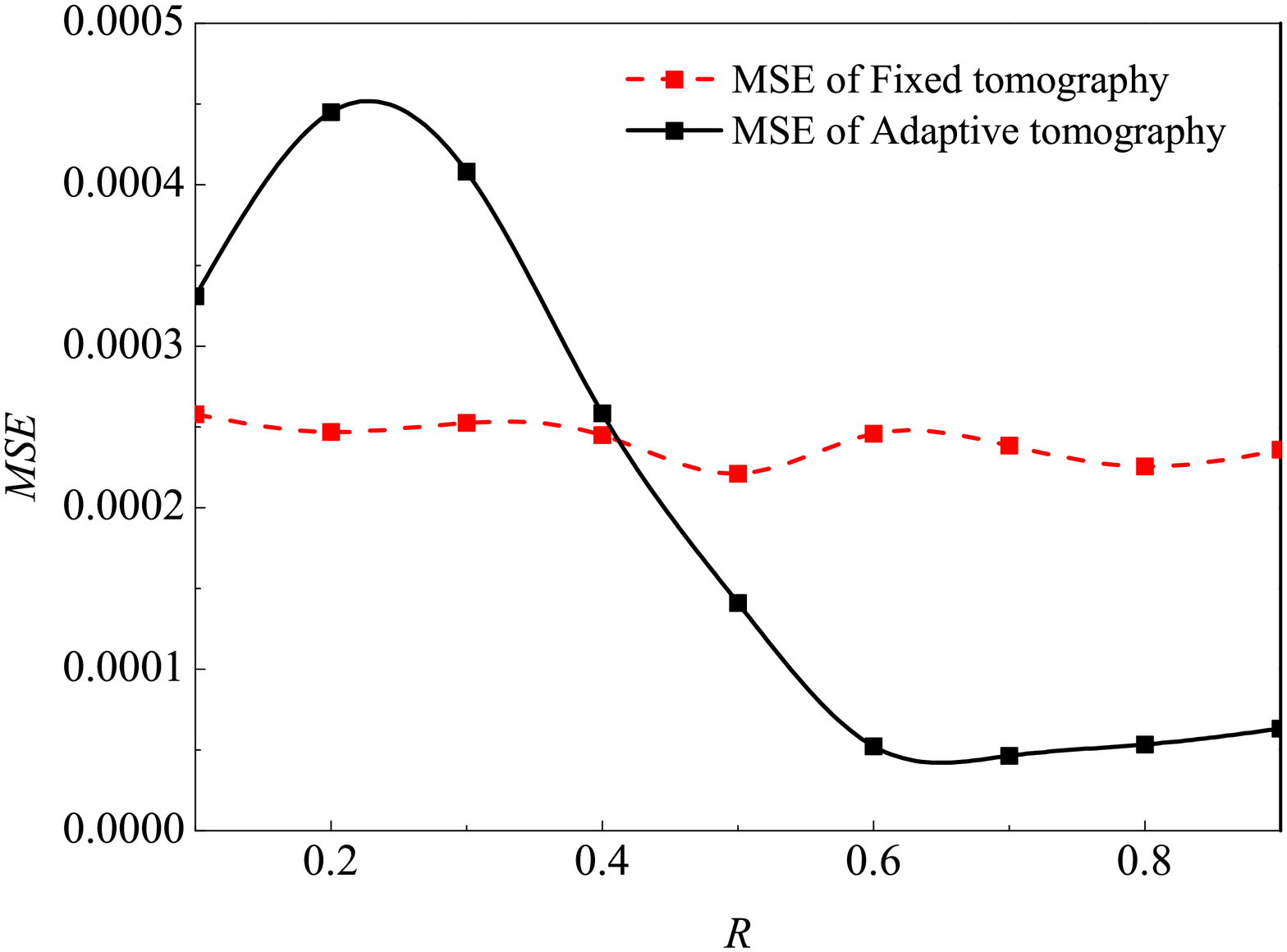}
     \caption{\label{Xstates} The picture of MSE under different ratios of $R$. The MSE is between the true two-qubit state density matrix $\rho_T$ (Fig.\ref{G9}) and the finial reconstructed density matrix $\rho_T$. It is calculated when same number of samples (90000) of $\rho_T$ is distributed under different $R$'s. Each point is gain by averaging 100 different values (MSE) under the same $R$.}
\end{figure}

When the $\rho_T$ is a randomly produced state (Fig.\ref{G9}), $90000$ copies of the state $\rho_T$ are applied for both fixed QST and AQST. For AQST, $90000\cdot R$ copies of the state $\rho_T$ are applied in the first round of measurement, while $90000(1-R)$ copies $\rho_T$ are applied in the second round of measurement. It is found that when the ratio $R$ is larger than 0.5, AQST produces a better result than fixed QST ($10^{-4}$ of MSE). From Fig.\ref{Xstates}, the optimal $R$ for AQST is nearly $2/3$ for this randomly produced state $\rho_T$. In other words, $N\cdot2/3$ samples of the state $\rho_T$ should be costed in the first step of the experiment, which confirms that the 1/3 used in (Ref.\cite{thePaperOfGuoLaoshi}) for $R$ may not be the optimal one.

In the following part, the three-step AQST is further implemented based on the previous steps to get a smaller value of MSE with the same number of copies of $\rho_T$. Based on the density matrix $\rho_{E1}$ obtained by the two-step AQST, the selection mechanism of the measurement setting is applied again. Then new frequencies can be obtained after using the same process as in the two-step AQST, and a more precise density matrix is estimated, as shown in Fig.\ref{shiyituerThreestep}. In the numerical test, the density matrix of Fig.\ref{G9} is applied and the $R$ is selected as $2/3$ according to Fig.\ref{Xstates}.

\begin{figure}
    \centering
    \includegraphics[width=10cm]{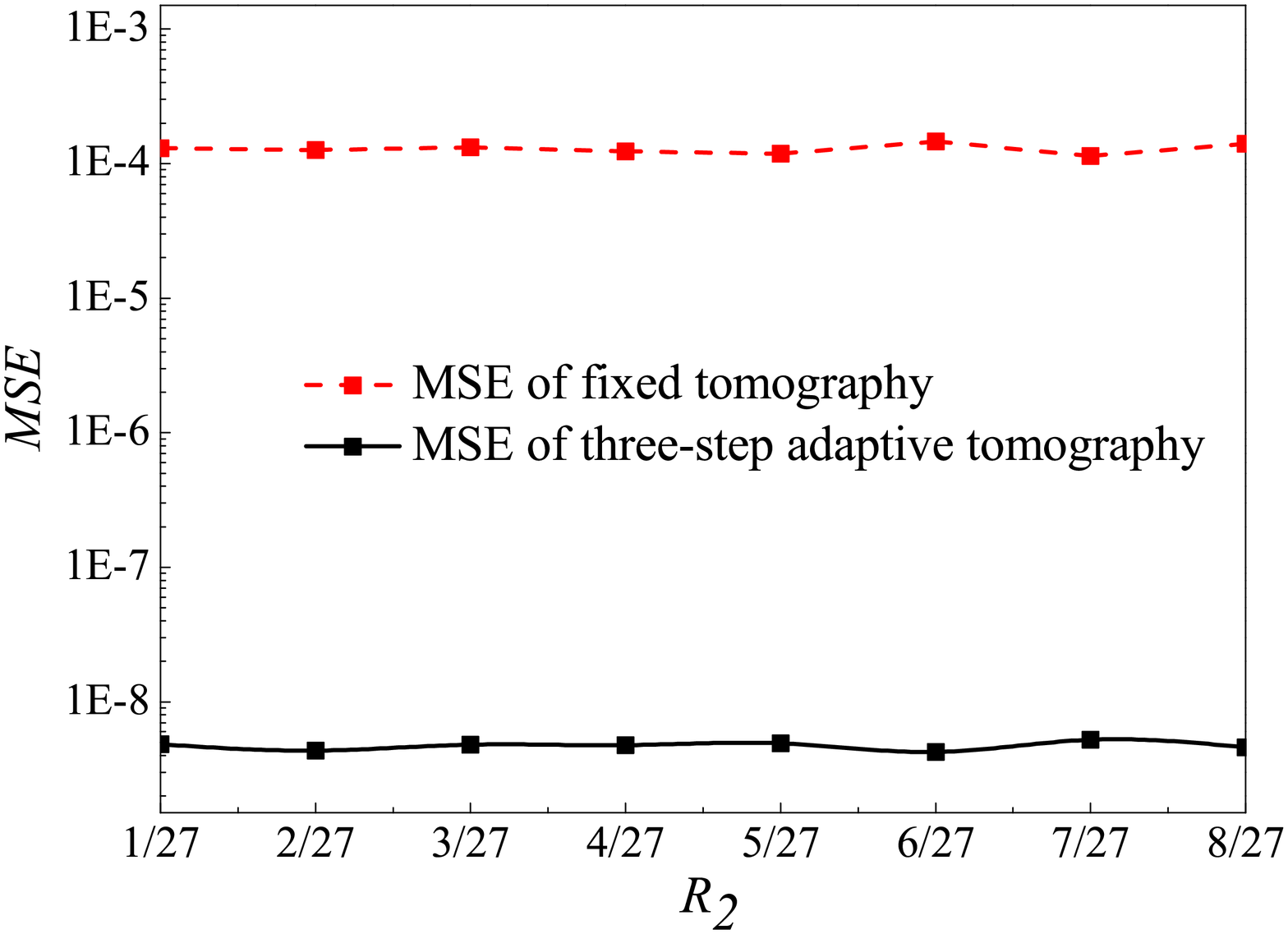}
    \caption{\label{ThreestepTomo} The picture of MSE under different $R_2$. Three-step AQST outperforms two-step AQST.}
\end{figure}

Then the test is performed according to the different number of copies of the  state $\rho_T$ in Fig.\ref{G9} distributed in the third step while the total number of copies are all equal. Define $R_2$ as the ratio of the number of copies of the state $\rho_T$ spent in the third step to the total number of copies of $\rho_T$ in QST. It is noticed that the three-step QST has greatly improved. In other words, the three-step AQST can give a better estimation of the density matrix under the same copies of the unknown state, as shown in Fig.\ref{ThreestepTomo}.

\section{Discussions}

 Frequency adjustment of Eq.(\ref{FUM}) is suitable to some special "sparse state" as is seen from the results of a large number of numerical simulations. In Fig.\ref{Sc} and Fig.\ref{G9}, the tested matrices are sparse. A sparse matrix means many zero entries in the density matrix, such as the pure SC state. It is also observed that the larger the gap between the modules of different elements in the true density matrix $\rho_T$, the better AQST performs under our mechanism of applying specific selection. The essence of the mechanism is to use the characteristics of the sparsity of the density matrix in the partial basis. Fixed QST is assigned to the same number of copies of $\rho_T$ on each measurement setting. Hence the prior information of the density matrix is not utilized. In contrast, AQST exploits the rough information about the  magnitude of each element of the preliminary density matrix, which is obtained using the partial copies of the unknown state $\rho_T$. The remaining samples of $\rho_T$ are better exploited because of the information provided by this rough density matrix.

The reason why we use AQST is briefly stated as follows:
The error of an entry of the density matrix becomes larger when the magnitude of the entry becomes larger but not exceeding 1/2 (and the cases of the magnitude exceeding 1/2 is very rare practically); this fact is obtained from Eq.(4) in Ref.\cite{D.H.Mahler}. This implies that the treatment of the remaining copies of the unknown state $\rho_T$ as distributed uniformly in the measurement settings is not the best choice. In order to reduce the big errors associated with the entries with large magnitude, we may distribute more copies of the unknown state $\rho_T$  in the settings that contain bases which the projected value of $\rho_T$ onto the bases is large. Then the total error in the density matrix, which is described by MSE, can be made smaller. In this paper, only MSE is utilized to compare the performance of AQST with that of the fixed tomography. More physical quantities, such as fidelity, are required to fully compare these methods. In addition, four-step AQST, or AQST with even more steps is not considered since the MSE provided by three-step is small enough to meet the requirement of QST and too many steps also means to reconstruct density matrix for many times, which will cost much time.

The purpose of allocation mechanism of the copy of a state is to obtain the density matrix as exactly as possible. Our results are especially favorable for multi-photon systems. The preparation of a large number of samples in short time is very challenging. Eight-photon system requires 170 hours to prepare a reasonable number of copies of the eight-photon SC state, which are just enough to certify the entanglement of these copies \cite{2}. Further characterization of the quantum state in experiment meets even more difficulties, and the current experimental technology is not up to the requirements.

In addition, it is also difficult to obtain the density matrix of large dimension by the inversion of relative frequencies obtained for POVMs according to the Born's rule in a short period of time. It will also take up a lot of storage space for POVMs in a computer. So based on the characteristics of POVMs, PhaseLift saves the huge storage space; the occupied space has been reduced from $\textit{O}(2^n\times2^n)$ to $\textit{O}(2^n)$. The multi-step AQST via PhaseLift may be extended to quantum process tomography under some conditions in the near future.

\section{Conclusion}

In AQST, a new scheme to distribute the copies of a state in different settings is developed based on the priori that some settings have more contributions to exactly estimate the density matrix. In addition, PhaseLift is applied to reconstruct the density matrix by saving the memory space required by the computer. By this new scheme with PhaseLift, a much more accurate density matrix can be obtained. The three-step AQST is also studied. By comparing their MSE, both the two-step and three-step AQST are shown to have a much better performance than the fixed QST.

\begin{acknowledgments}
 The authors would like to thank Prof. Jian-wei Pan, Prof. Chaoyang Lu and other members of their group. The authors also would like to thank Dr.Yulong Liu and Dr.Wenkai Yu for helpful discussions. The authors would like to greatly thank Prof. Mo-Lin Ge for the valuable discussions. We further express our sincere gratitude to Prof. K. Fujikawa for making revisions of article. This work is supported by NSF of China with the Grant No. 11275024 and No.11475088. Additional support was provided by the Ministry of Science and Technology of China (2013YQ030595-3, and 2013AA122901).
\end{acknowledgments}


%
%

%



\end{document}